\documentclass{aa}
\usepackage{graphicx}
\usepackage{natbib}
\newcommand{\etal}{{\em et~al.}}
\begin{document}

\title{Gamma photometric redshifts for long gamma-ray bursts}

\author{Z. Bagoly \inst{1} \and I. Csabai \inst{2}
\and A. M\'esz\'aros \inst{3} \and P. M\'esz\'aros \inst{4}
\and I. Horv\'ath \inst{5} \and L. G. Bal\'azs \inst{6}
\and R. Vavrek \inst{7}}

\offprints{Z. Bagoly} 

\institute{Laboratory for Information Technology, E\"{o}tv\"{o}s
 University, H-1117 Budapest, P\'azm\'any P. s.  1./A, Hungary\\
 \email{bagoly@ludens.elte.hu}
 \and 
 Dept. of Physics for Complex Systems,
E\"{o}tv\"{o}s University, H-1117 Budapest, P\'azm\'any P. s.  1./A, Hungary\\
 \email{csabai@ludens.elte.hu}
 \and 
 Astronomical Institute of the Charles University,
   V Hole\v{s}ovi\v{c}k\'ach 2, CZ-180 00 Prague 8, Czech Republic\\
\email{meszaros@mbox.cesnet.cz}
\and 
  Dept. of  Astronomy \& Astrophysics, Pennsylvania State University,
  525 Davey Lab., University Park, PA 16802, USA\\
\email{nnp@astro.psu.edu}
\and
 Dept. of Physics, Bolyai Military University, H-1456 Budapest, POB 12, 
Hungary\\ \email{hoi@bjkmf.hu}
\and 
 Konkoly Observatory, H-1505 Budapest, POB 67, Hungary\\
\email{balazs@ogyalla.konkoly.hu}
\and 
Max-Planck-Institut f\"{u}r Astronomie, D-69117 Heidelberg, 17 K\"{o}nigstuhl,
Germany \\  \email{vavrek@mpia-hd.mpg.de}}

\date{Received 18 June, 2002/Accepted 18 November 2002 }

\abstract{
It is known that the soft tail of the gamma-ray bursts' spectra show
excesses from the exact power-law dependence. In this article we show
that this departure can be detected in the peak flux ratios of
different BATSE DISCSC energy channels. This effect allows to
estimate the redshift of the 
bright long gamma-ray bursts in the
BATSE Catalog. A verification of these redshifts is obtained 
for the 8 GRB which have both BATSE DISCSC data and measured optical
spectroscopic redshifts. There is good correlation between the measured and estimated
redshifts, and the average error is $\Delta z \approx 0.33$.
The method is similar to the photometric redshift
estimation of galaxies in the optical range, hence it can be
called as "gamma photometric redshift estimation". The estimated redshifts
for the long bright gamma-ray bursts are up to $z \simeq 4$. For the
the faint long bursts - which should be up to $z \simeq 20$ - the redshifts
cannot be determined unambiguously with this method.
                    }
		 
\maketitle
		 
\section{Introduction}

The gamma-ray bursts (hereafter GRBs) of the long subgroup
\citep{Kouveliotouetal1993} detected by the 
BATSE instrument \citep{Meeganetal2000}
are at high redshifts. The highest directly measured redshift is at
$z= 4.5$ \citep{Andersenetal2000,Mesz2001}, but there are indirect
considerations - based on BATSE data - predicting the existence of redshifts
up to $z \simeq 20$ \citep{MeszarosMeszaros1995,
MeszarosMeszaros1996, HorvathMeszarosMeszaros1996,ba98}.
This result is based on distribution densities and deals with 
the GRB redshifts in statistical sense only. This means that
one may obtain the fraction of GRBs being at a given redshift interval
(see, for example, \citet{Schmidt2001}), but one cannot obtain the redshift of a 
given GRB event.

There are only a few cases, when
the observation with the BeppoSAX satellite \citep{Piroetal2002} or
other instruments \citep{Klose2000} made possible to detect the afterglows
and then the measurement of redshifts using optical spectroscopy.
The Current BATSE Catalog \citep{Meeganetal2000} 
consists of more than $1200$ long bursts,
but for only $9$ of them have redshift measurement (8 have redshifts 
and there is one GRB with an upper redshift limit). The unfortunate
premature termination of the CGRO satellite prevents to increase this
number further. There are other instruments observing $\approx 100$
bursts/year, but the typical number of burst's redshifts is only about a
dozen/year \citep{Mesz2001}. 

Hence, any method that could estimate the 
redshifts from X-ray/gamma-ray observations 
{\it alone} would be a great help.

In \citet{Ramirez-RuizFenimore2000} and \citet{Reichartetal2001} a linear
relation between the intrinsic peak-luminosities of GRBs and their so
called "variabilities" was found.
Similarly, \citet{Norrisetal2000a} found a relation between the so called
spectral lag and the peak-luminosity allowing to estimate the redshifts
of long GRBs. These relations were calibrated on a few
cases of GRBs, when GRBs were observed {\it both} by BATSE and
other instruments measuring the optical redshift from afterglows.
Then, having either the variabilities or the spectral lag of a given
GRB, one can estimate its redshift. 
The physical meaning of the correlation between the
variability (spectral lag) and the peak-luminosity remains unclear.
These two methods can be combined \citep{Schaeferetal2001} to determine
redshifts if all the needed input parameters are available for the GRBs.

In this article we present a new method of the estimation of the
redshifts for the long GRBs.
The situation is in some sense similar to the optical observations of
galaxies, where the number of objects with broad band photometric
observations is much larger than the number of objects with measured
spectroscopic redshifts. For galaxies and quasars the growing field of 
photometric redshift estimation \citep{koo85, connolly95a, gwyn96,
sawicki97, wang98, soto99, benitez00, csabai00, budavari00, 
budavari01} achieved a great success in estimating redshifts from photometry 
only. 
Here we present a method that is
quite similar to these methods; hence we call it as {\it gamma
photometric redshift estimation} (GPZ for short). We utilize the fact
that broadband fluxes change systematically, as characteristic spectral
features redshift into, or out of the observational bands. 
Hence, contrary to the variability and spectral lag methods, 
this technique has a well defined physical meaning. 

The article is structured as follows.
First, using a spectral model for GRBs  we deduce 
an expected relation between a measurable quantity (peak flux ratio) 
and the redshift (Sect. 2). Having this relation, we verify it on the existing
sample of a few GRBs having measured redshifts (Sect. 3). 
Because both Sect. 2 and Sect. 3 suggest that this method
is usable, Sect. 4 presents the estimated redshifts for hundreds of
long GRBs. In Sect. 5 we discuss and summarize the results.

\section{Gamma photometric redshift estimation}

To understand our method in this Section we
outline the general scheme of 
broadband observations. The method is generally the same both for the optical
and gamma-ray ranges. 
The only major difference is that in the X-ray and gamma-ray range the 
extra- and intergalactic medium have negligible effects, but the optical 
photons are attenuated. 

Let us take two different instrumental channels defined by 
$E_4 > E_3$ and  $E_2 > E_1$. 
If one would know the rest-frame energy spectrum ($L(E)$) of the
burst, for a perfect instrument that captures all the photons in the
above energy channels, the observed luminosity (in units photons/s) 
for a burst at redshift $z$ would be the following:

\begin{equation}
L_{2,1} = \int_{(1+z)E_1}^{(1+z)E_2} L(E) dE, \;\;
L_{4,3} = \int_{(1+z)E_3}^{(1+z)E_4} L(E) dE,
\end{equation}

For the observed fluxes ($P$) similar equations can be used. Let us
define the following flux ratio:

\begin{equation}
R(E_4, E_3, E_2, E_1, z) = \frac{L_{4,3} - L_{2,1}}{L_{4,3} + L_{2,1}} = \frac{P_{4,3} - P_{2,1}}{P_{4,3} + P_{2,1}}
\end{equation}
which in general depends on the redshift, since $L_{2,1}$ and $L_{4,3}$
depends on the redshift.

Assume for the moment that one observes a pure power-law spectrum. This
means that
$L_E \propto E^{-\alpha}$ holds, where the exponent $\alpha$ is a real
number. In this special case one could prove easily that for any
redshift $z$
\begin{equation}
\frac{{L}^{z=0}_{4,3} - {L}^{z=0}_{2,1}}
{{L}^{z=0}_{4,3} + {L}^{z=0}_{2,1}} =
\frac{L_{4,3} - L_{2,1}}{L_{4,3} + L_{2,1}},
\end{equation}
where
\begin{equation}
{L}^{z=0}_{2,1} = \int_{E_1}^{E_2} L(E) \; dE, \;\;
{L}^{z=0}_{4,3} = \int_{E_3}^{E_4} L(E) \; dE.
\end{equation}
This means that in this special case $R =R(E_4, E_3, E_2, E_1, z)$
is {\it not} depending on $z$.

Of course, in the real situation, the spectrum has got a more complicated
form, and hence $R$ is depending on $z$; this will be the effect that we
will use for redshift estimation.

In addition, in the real situation, 
the incident spectrum measured by the detector is 
convolved with the detector's response function defined by response
matrix \citep{Pendletonetal1994} resulting the measured flux
of the corresponding channel. For the channel with
energy range $E_2>E>E_1$ the measured flux $P_{1,2}$ is therefore given by
\begin{equation}
P_{1,2} = \int_{E_1}^{E_2} P(E) c(E) \; dE,
\end{equation}
where  $c(E)$ is the detector's response function. 
The similar holds for the second channel, too, 
with the same $c(E)$. Hence, in general, $R$ is depending 
both on the spectrum and the response function.

If the rest-frame spectrum for a GRB is known, one is able to 
calculate the theoretical $R$ as a function on $z$. Then these values can 
be compared with the flux ratio obtained from the broadband measurements
($R_{meas}$). The redshift, where $(R-R_{meas})^2$ is minimal, could give
the estimated gamma photometric redshift.

Regarding this gamma photometric redshift estimation, the major problem
comes from the fact that the spectra are changing quite rapidly with 
time; the typical timescale for the time variation is $\simeq (0.5 - 2.5)$~s
\citep{RydeSvensson1999,RydeSvensson2000}.
Hence, if possible, one should consider spectra
which are defined for time intervals smaller than this characteristic time.
Therefore, we will consider the spectra in the $320$~ms  time interval
(i.e. in five $64$~ms  time intervals), 
with the peak-flux being at the center of this time interval.

In the following we will {\em assume} that the spectrum has the same shape
around the time of the peak-flux for all long bursts. Unfortunately we
do not have any deep theoretical or observational evidence for this
assumption, instead we will
test our assumption on GRBs, where spectroscopic redshifts are available
(next Section). Because, this assumption seems to be acceptable, in Sect.
4 we will use $R$ to estimate $z$ for long GRBs. 

\section{Application on GRBs: Calibration}

It is well-known \citep{Bandetal1994,Amatietal2002} 
that the time-integrated average
spectra of GRBs can be approximated by a broken power-law; the
break is at some energy $E_o$.  
The typical {\em rest-frame} energy for $E_o$ is
above $\simeq 500$ keV \citep{Preeceetal2000a,Preeceetal2000b},
but this might vary for different GRBs.

Of course this broken power-law spectrum
is simply an approximation: first, because the break around $E_o$ may have a
more complicated form \citep{Preeceetal2000a,Preeceetal2000b},
and, second, because at low {\em rest-frame} energies (around $\simeq
80$ keV) there may be essential departures from the power-law.  This
is the so called soft-excess, which is confirmed for $\simeq 15\%$ of
GRBs on the high confidence level 
\citep{Preeceetal1996, Preeceetal2000a,Preeceetal2000b}; 
and for the remaining GRBs the
soft-excess seems to occur, too \citep{Preeceetal1996}.

Based on this, we construct our {\em template spectrum} that will be
used in the GPZ process in the following manner: Let the spectrum be a sum
of the Band's function \citep{Bandetal1994},
and of a low energy power-law function taking the form
\begin{eqnarray}
\mbox{for}\;  E \leq E_o  & L(E) & = a (E/E_{cr})^{-\alpha} + a 
(E/E_{cr})^{-\beta}  \\
\mbox{for}\;  E \geq E_o  & L(E) & = a_3 (E/E_o)^{-\gamma}, 
\label{template:spectrum}
\end{eqnarray} 
where $a_3 = a [(E_o/E_{cr})^{-\alpha} + (E_o/E_{cr})^{-\beta}] $
comes from the normalization.
In this spectrum there are six parameters, but the amplitude $a$ is for $R$
unimportant, and need not be specified.
We fix all the above parameters according to the available literature,
so there are no free tunable parameters in our method. 
The low energy cross-over is at $E_{cr} 
= 90$ keV,  $E_o=500$ keV, and the spectral indices are
$\alpha=3.2$, $\beta=0.5$ and $\gamma=3.0$  
\citep{Preeceetal2000a, Preeceetal2000b}.

As we remarked above, the
spectrum is rapidly changing. But our assumption is that the spectrum has a
characteristic shape around the instant of the peak. 
Now we have to chose a short time interval around the peak (maximum of the total counts), 
during which the change of spectrum is still negligible, but the number
of photons allows good signal-to-noise ratio. To be able to
cut out such time interval around the peak-flux, we need data with
a reasonably good time resolution. In our study we will use the $64$~ms 
resolution BATSE LAD DISCSC data from the public BATSE
Catalog \citep{Meeganetal2000}. During this time interval
the change of our template spectrum is still negligible \citep{RydeSvensson2000}.

We have also checked the robustness of the PFR against the 
integration time around the peak. Both the doubling of the integration
time (for $640$~ms) and its skewing around the peak 
did not change significantly the values of PFR. All this means that
the template spectrum defined by Eqs.(6-7) seems to be a good approximation
for the $320$~ms  time interval around the peak. This is in fact
expectable from earlier studies of spectra \citep{Bandetal1994,
RydeSvensson1999,RydeSvensson2000}.

\begin{figure}
\begin{center}
\includegraphics[width=8.2cm]{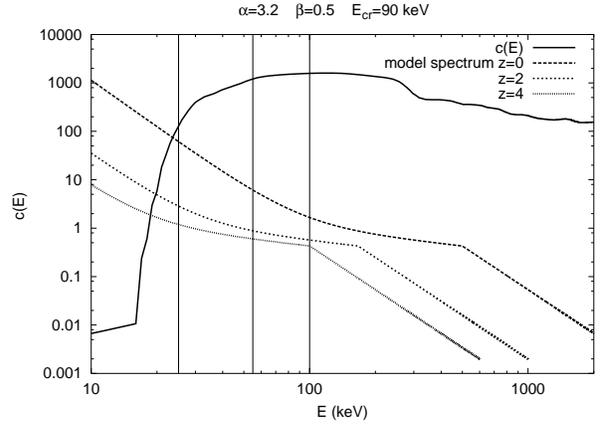}
\caption{The detector response function $c(E)$
and the behaviour of a template spectrum at different redshifts.}
\label{fig:response}
\end{center}
\end{figure}

 The 4 energies for the BATSE instrument are: 
$E_1=25$ keV, $E_2=E_3=55$ keV, $E_4=100$ keV. 
Using the detector response matrices \citep{Pendletonetal1994}
one can calculate the observed counts and flux for any incoming spectrum.
In Fig.~\ref{fig:response} we show a typical response
function $c(E)$. The response function is different for each burst,
but using the BATSE DRM data one can use the actual response
function for every burst. Fig.~\ref{fig:response} also demonstrates the
behaviour of the spectrum at different redshifts. Going from
$z=0$ to higher redshifts one can see that the soft-excess moves from
the second channel to the first one and then leaves the range of this
detector around $z\approx 4$.

\begin{figure}
\begin{center}
\includegraphics[width=8.2cm]{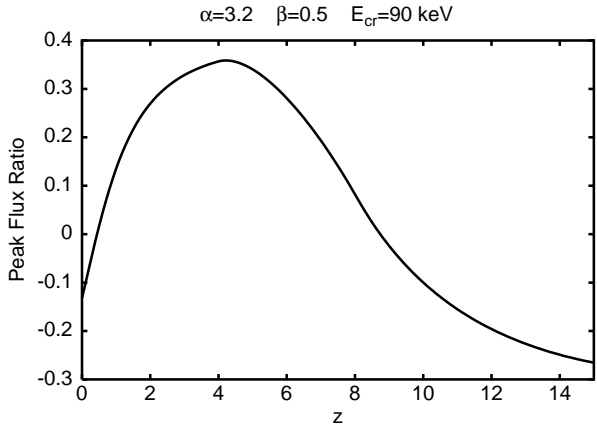}
\end{center}
\caption{The theoretical PFR curves 
calculated from the template spectrum using the 
average detector response matrix.} 
\label{fig:serTheor}
\end{figure}

Before starting the detailed investigation of the fluxes that one can
get using the template spectra and the response matrices
by Eq.(6), let us test the 
correctness of the template spectrum in a simple way. Let us 
introduce the {\em peak flux ratio} (PFR hereafter) in the following way:
\begin{equation}
\mbox{PFR} = {{l_{34}-l_{12}} \over {l_{34}+l_{12}}}
\end{equation}
where $l_{ij}$ is the BATSE DISCSC flux in energy channel $E_i < E < E_j$
integrated for $320$~ms  around the peak flux.
(I.e. five $64$~ms  intervals are summed - the middle one is where the flux
is the biggest. Even during this time interval the change of spectrum is still
negligible \citep{RydeSvensson2000}). Our theory says that for the 
above template spectrum this ratio should increase with $z$.
On Fig.~\ref{fig:serTheor} we plot the theoretical
PFR curves calculated from the above defined template spectrum using the 
{\em average} detector response matrices for the 9 bursts that have both BATSE
data and measured redshifts. These bursts' data are collected in Table 1. 

\begin{table}
\caption{The redshifts of GRBs that have both BATSE
triggers and measured spectroscopic redshifts. 
Data compiled by \citet{Klose2000}; see also \citet{Bloometal2001}.}
$$
\begin{array}{lllll}
\hline     
\mbox{Burst} & \mbox{BATSE}   & z & &\mbox{Remark}\\
       & \mbox{trigger} &   & & \\
\hline
980425 & 6707 & 0.00857 & & \mbox{SN1998bw} \\
970508 & 6225 & 0.8356 & & \\
970828 & 6350 & 0.9578 & & \mbox{no DISCSC data} \\
980703 & 6891 & 0.9676 & & \\
991216 & 7906 & 1.020 & & \\
990123 & 7343 & 1.6006 & & \\
990510 & 7560 & 1.6196 & & \\
971214 & 6533 & 3.4127 &  & \\
980329 & 6665 & 3.5 & & \mbox{upper limit only}\\
 \hline
   \end{array}
$$
   \label{table1}
   \end{table}
   
\begin{figure}
\begin{center}
\includegraphics[width=8.2cm]{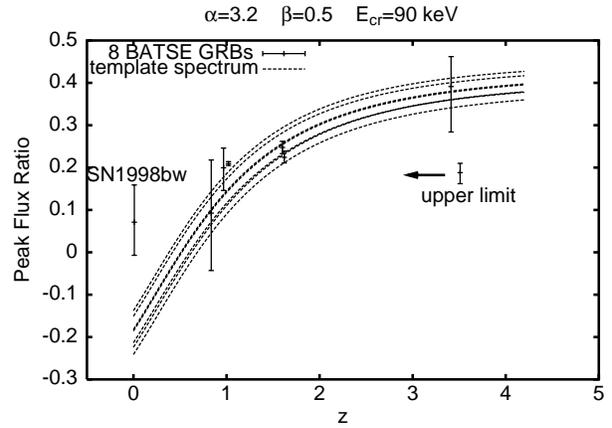}
\end{center}
\caption{Experimental (points with error bars) and theoretical (dashed
lines) peak flux ratio values for the 8 bursts that have both BATSE DISCSC 
data and measured redshifts.}
\label{fig:serz} 
\end{figure}

If we redshift the template spectrum and apply the corresponding
detector response matrix of the given burst, we can get a PFR value
for any redshift. Fig.~\ref{fig:serz} shows the theoretical
curves together with the seven PFR values calculated from  observed GRB 
data (the DISCSC data for GRB 970828 are missing). 
We used the template spectrum defined by Eqs.(6-7) 
and the different response matrices 
corresponding to the observational conditions of the given burst. 
There is a clear trend: as expected from the above
considerations, PFR increases with increasing redshifts up to $z \simeq 4$.
Except for the GRB associated with the supernova, 
and for the GRB having only the upper limit,
the remaining 6 GRBs have a clearly increasing PFR with increasing $z$. 

\begin{figure}
\begin{center}
\includegraphics[width=8.2cm]{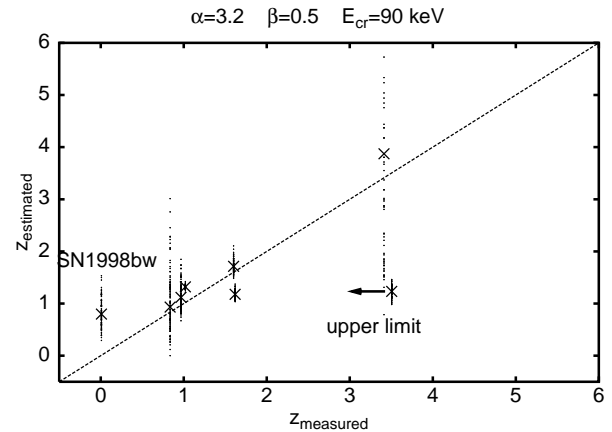}
\end{center}
\caption{The measured spectroscopic redshift values compared 
with gamma photometric redshift estimation for the 8 considered GRBs.}
 \label{fig:zz} 
 \end{figure}

In the used range of $z$ (i.e. for $z ^{<}_{\sim} 4$)
 the relation between $z$ and PFR is
invertable. Hence we can use it to estimate the 
{\em gamma photometric
redshift} (GPZ) from a measured PFR. In Fig.~\ref{fig:zz} the measured
spectroscopic redshifts are compared with GPZ 
values for 8 considered GRBs. 
The errorbars show the effect of counts' Poisson noise only.

Leaving out GRB associated with the supernova 
and GRB having upper redshift limit only, 
the estimation error is
$\Delta z=\sqrt{\sum_{i=1}^6 (z^{spec}_i-z^{GPZ}_i)^2/5} \approx 0.33$.

In order to test the reality of the correlation between  the soft excess
and the  redshift we made the null hypothesis that there is no
relationship between these quantities, i.e the computed correlation is
purely random. Assuming no true correlation between the soft excess and
redshift the probability density of the computed
quantity can be given by

$$ f(x)=\frac{1}{\sqrt \pi
}\frac{\Gamma((N-1)/2)}{\Gamma((N-2)/2)}(1-x^2)^{(N-4)/2} ,$$\\
where $N$ is number of data points \citep{Spiegel1999}. Let $r_c$ be the
calculated correlation. Then the $\beta$ error probability at rejecting the null
hypothesis is given by

$$\beta=1- \int \limits_{-1}^{r_c} f(x) dx .$$

In Table 2. the $N$, $r_c$ and the  calculated level of significance are shown 
for various cases.

\begin{table}
\caption{The linear correlation coefficients between $z^{spec}$ and $z^{GPZ}$ }
$$
\begin{array}{crrc}
\hline     
\multicolumn{1}{c}{N}     &     \multicolumn{1}{c}{r_c}    &      \multicolumn{1}{c}{p = 1-\beta}  &        \multicolumn{1}{c}{\mbox{Remark}}	\\
\hline
    7     &    0.9364 &   0.9991 &	\\
    6     &    0.9655 &   0.9991 &      \mbox{GRB 980425 excluded}\\
    6     &    0.7809 &   0.9666 &      \mbox{GRB 971214 excluded}\\
    5     &    0.6618 &   0.8881 &      \mbox{GRB 980425 + GRB 971214 excluded}\\
\hline
\end{array}
$$
   \label{table2}
   \end{table}

Although it seems that $z^{GPZ}$ for GRB 971214 (where $z=3.4127$)
fits very well the estimation error without it is better: $\Delta z \approx
0.29$. However the linear correlation coefficient here with $N=5$ yields a 
much poorer $r_c=0.66$ with a $p=0.89$ significance.

We see that PFR (if calculable from observations for the given burst) is a 
quantity that may allow to determine  redshift.
Problems may arise from the fact that for any value of PFR two redshifts are
possible - either below or above $z \simeq 4$ (see Fig.~\ref{fig:serTheor}),
further measurements are needed to exclude one of the redshift.

\section{Application on GRBs: Estimation of the redshifts}

To avoid the problems with the instrumental threshold
we exclude the faintest GRBs from the BATSE data.
Similarly to \citet{pend} and \citet{ba98} these 
events have a $F_{256}$ peak-flux (i.e. on $256$~ms  trigger scale) 
smaller than $0.65$ photon/(cm${}^2$s). These GRBs are not discussed in this
article.

Further restriction comes from the fact that short GRBs are today taken
as different phenomena \citep{Horvathetal2000,Norrisetal2000b}. 
In addition, due to instrumental effects \citep{Piroetal2002}, 
no spectroscopic redshifts are known for this subgroup of GRBs. Hence,
we do not apply our method for short GRBs.

The reality of the
intermediate subgroup of GRBs \citep{Horvath1998, Mukherjee1998, HA,  br, rm, ho2}
having remarkable sky angular distribution
\citep{Meszarosetal2000a, Meszarosetal2000b, li01} 
is unclear yet. 
In any case, no spectroscopic redshifts are known also here. Hence,
we exclude this subgroup, too.

Therefore, we restrict ourselves to long GRBs defined by $T_{90}$ 
\citep{Meeganetal2000} with $T_{90} > 10\;s$ . 
There are 1241 GRBs in BATSE Catalog fulfilling
this condition. Deleting GRBs having no $F_{256}$ and having
$F_{256} < 0.65$ photon/(cm${}^2$s) 838 GRBs remain. This sample is studied
here.

Introducing an another cut $F_{256} > 2.00$ photon/(cm${}^2$s) 
we can investigate roughly the brighter half of this sample.
We will discuss the sample $F_{256} > 2.00$  photon/(cm${}^2$s)
("bright half" sample having 343 GRBs) and
$F_{256} > 0.65$ photon/(cm${}^2$s) 
("all" sample having 838 GRBs), respectively.

As the soft-excess range redshifts out from the BATSE DISCSC energy
channels around $z \approx 4$, the theoretical curves converge to a
constant value. For higher $z$ it starts to decrease. 
This is where the power-law breakpoint
($E_o$) is redshifts into soft energy range. This means that the
method is ambiguous: for the given value of PFR one may have two
redshifts - below and above $z \approx 4$. Because for
the bright GRBs the values above $z \approx 4$ are practically excluded,
for them the method is usable. In other words,
using only the $25-55$ keV and $55-100$ keV BATSE energy channels, 
this method can be used to estimate GPZ only in the redshift range 
$z\; ^{<}_{\sim}\; 4$; outside of this region the $z$
vs. PFR relation is non-invertable (see Fig.~\ref{fig:serTheor}).
For high redshifts (above $z \approx 4$) 
the method gives two possible values. For faint GRBs the estimation also usable
(at least in principle), but one has to decide by other arguments that either
the redshift below $z \approx 4$ or above $z \approx 4$ is the correct value.

\begin{figure}
\begin{center}
\includegraphics[width=8.2cm]{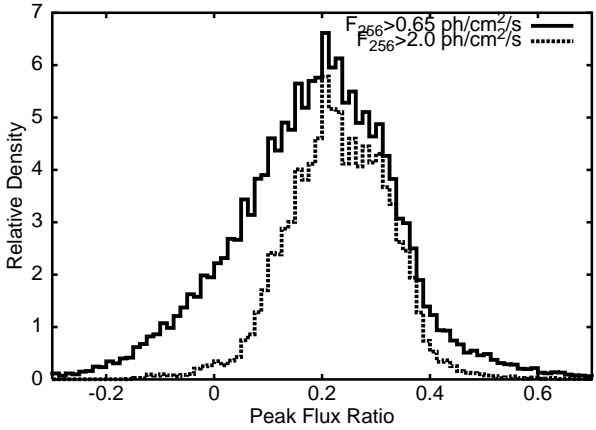}
\end{center}
\caption{The PFR distribution of the long GRBs having DISCSC data.} 
\label{fig:PFRDist} 
\end{figure}

Let us assume for a moment that all observed long bursts, we have
selected above,  have $z < 4$. 
Then we can simply calculate the $z^{GPZ}$ redshift 
for any GRB, which has calculable PFR from BATSE DISCSC data.
Fig.~\ref{fig:PFRDist} shows the distribution of 
the measured PFRs of the long GRBs having DISCSC data.
The fact that the number of objects beyond the minimal and maximal
theoretical PFR values ($\approx -0.15$ and $\approx 0.37$,
respectively) is relatively small, is reassuring. 

\begin{figure}
\begin{center}
\includegraphics[width=8.2cm]{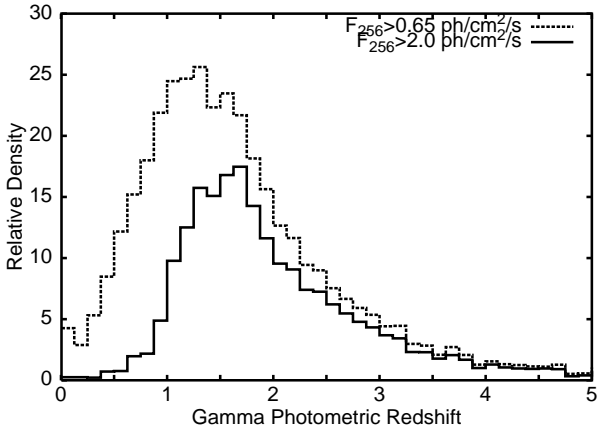}
\end{center}
\caption{The distribution of the Gamma Photometric Redshift estimators 
of the long GRBs having DISCSC data. The distribution of the 
bright half of the BATSE Catalog is also shown.}
\label{fig:zDist} 
\end{figure}

Fig.~\ref{fig:zDist} shows the distribution of the estimated 
derived redshifts {\it under the assumption that all GRBs are below} $z
\approx 4$. 
The distribution has a clear peak value around PFR $\approx 0.2$, which
corresponds to $z \approx (1.5- 2.0)$.

\section{Discussion}

Having the estimated redshifts shown on Fig.~\ref{fig:zDist} one may ask:
Are these redshifts really correct? 

There can be two different problems here. First of all,
the method is based on the assumption that around the peak flux, the
spectrum is the same for all the selected long GRBs. Second, the method 
gives degenerate result,  with two possible redshift values.

Concerning the first problem we could just hope that in the near future
some theoretical or experimental evidence will confirm our assumption, 
but the situation is not worse than
in the studies of \citet{Norrisetal2000a} and \citet{Reichartetal2001}. These
articles also suggest that despite the deeper understanding of the
underlying physics, the procedure itself is usable. 
In addition, here the PFR-$z$ 
relation is well supported by earlier independent observations.

Concerning the second problem we think that the great majority 
of values of $z$ obtained for the bright half are correct. 
 This opinion may be supported by three independent arguments. First,
the obtained distribution of GRBs 
in $z$ for the bright half on Fig.~\ref{fig:zDist}
is very similar to the obtained distribution of \citet{Schmidt2001}
(see Fig.~6 of that article). The luminosity-based redshift distribution 
\citep{Schaeferetal2001} also suggest an uniformly rising GRB density out to 
$z \approx 5$.  Second, as $z$ 
moves into $z ^{>}_{\sim} 4$ regime for the bright GRB, one would obtain
extremely high luminosities. Using Eq.12 of 
\citet{MeszarosMeszaros1996}, there is a {\it lower limit} for the isotropic
luminosity of the GRBs a value $\simeq 10^{53} ergs/s$. 
(Note here that the precise value is, of course, calculable
and is depending on the chosen cosmology model and on the typical 
energy of emitted photons. For the purpose of this article this approximate
value is enough.) This is an unacceptable high value for a 
lower limit, because typical luminosities are $\simeq 10^{51-52} ergs/s$. 
(see, e.g., Table 1 of \citet{Reichartetal2001}). We cannot exclude that a few
cases from bright GRB are at $z ^{>}_{\sim} 4$, but - we think - in the bright
half this cases are rare.  
Thirds, as an additional statistical test we compared the redshift distribution of 
the 17 GRB with observed redshift with our reconstructed GRB $z$ distributions 
(limited to the $z<4$ range).  For the  $F_{256} > 0.65$ photon/(cm${}^2$s) group 
the Kolmogorov-Smirnov test suggests a $38\%$ probability, i.e. the observed 
$N(<z)$ probability distribution agrees quite well with the GPZ reconstructed 
function. Although the observed distribution suffers 
from strong selection effects this fact is nevertheless reassuring.

\begin{figure}
\begin{center}
\includegraphics[width=8.2cm]{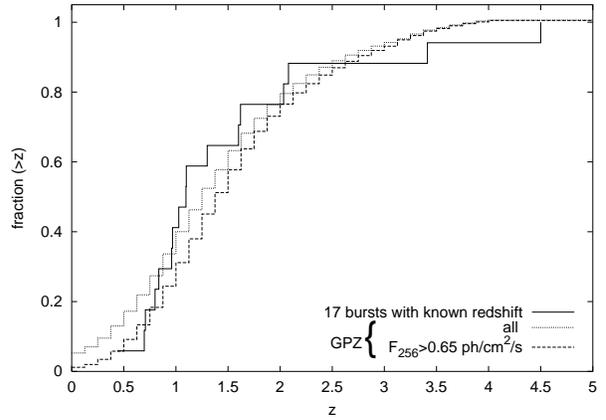}
\end{center}
\caption{
The redshift distribution of the 17 GRBs' with known redshift
and the distributions from the Gamma Photometric Redshift estimators 
for different peak flux groups.  }
\label{fig:KS} 
\end{figure}

For the faint GRBs being between
$F_{256} = 0.65$ photon/(cm${}^2$s) and $F_{256} = 2.00$ photon/(cm${}^2$s)
the situation is different. 
From Fig.~\ref{fig:zDist}
it follows that for $z < 1.7$ GRBs should be dominated by faint objects.
From this Figure one would obtain 
that GRBs are in average at smaller redshifts. This is
clearly a wrong conclusion, which is caused by the false assumption
that $z < 4$ for all the faint GRBs. We think that the
majority of faint GRBs $z$ should be changed into the value $z ^{>}_{\sim} 4$. 
Unfortunately, we are not able to say, concretely which GRB has a great
($z ^{>}_{\sim} 4$), and which GRB has still a small ($z ^{<}_{\sim} 4$)
redshift. 
In addition, for these faint GRBs also the error of estimated redshifts 
should probably be bigger than $\Delta z \approx 0.33$.
Simply, we conclude that in the current form with the current data, our method 
is not applicable for the faint GRBs.

The results of this work may be summarized as follows.
\begin{enumerate}
\item Based on earlier observations of GRB spectra
it is shown that the {\em peak flux ratio (PFR)} should be a well defined 
function of $z$.
\item The estimated redshifts from PFR are in good accordance with 
the known redshifts of the few
GRBs in BATSE Catalog having spectroscopic redshifts
\item All this allows us to calculate the redshifts of long GRBs.
Unfortunately, due to the twofold character of the PFR curve, the method is
usable only for bright GRBs.
\item Redshift distribution of 343 bright long GRBs are determined (Fig.~\ref{fig:zDist}).
\end{enumerate} 

\begin{acknowledgements}
The useful remarks with Drs. T. Budav\'ari, S. Klose, D. Reichart, A.S. 
Szalay and the anonymous referees are kindly acknowledged. 
This research was supported in part through
OTKA grants T024027 (L.G.B.), F029461 (I.H.) and T034549,
Czech Research Grant J13/98: 113200004 (A.M.), NASA grant NAG5-9192 (P.M.). 
\end{acknowledgements}

  \end{document}